\documentclass[prd,12pt]{revtex4}
\usepackage{amsmath,amssymb} % Enhanced mathematics
\usepackage[dvips]{graphicx} % Include figure files
\usepackage{graphics}
\usepackage{graphicx}
\usepackage{epsfig}
\usepackage[english]{babel}
   \newcommand*{\Li}{\ensuremath{\mathrm{Li}_2}}

\newcommand\nn{\nonumber}
\newcommand\ba{\begin{eqnarray}}
\newcommand\ea{\end{eqnarray}}
\newcommand\ee{\end{equation}}
\newcommand\be{\begin{equation}}

\begin{document}

\title{Radiative proton-antiproton annihilation to a lepton pair}

\author{A.~I.~Ahmadov$^{1,2}$\footnote {E-mail: ahmadov@theor.jinr.ru}, V.~V.~Bytev$^1$\footnote {E-mail: bvv@jinr.ru}, E.~A.~Kuraev$^1$\footnote {E-mail: kuraev@theor.jinr.ru}, E.~Tomasi-Gustafsson$^3$\footnote {E-mail: egle.tomasi@cea.fr}}
\affiliation{$^1$ Joint Institute for Nuclear Research, Dubna, Russia}
\affiliation{$^2$ Institute of Physics, Azerbaijan National Academy of Sciences, Baku, Azerbaijan}
%\author{E.~Tomasi-Gustafsson}
\affiliation{$^3$ \it CEA,IRFU,SPhN, Saclay, 91191 Gif-sur-Yvette Cedex, France, and \\
CNRS/IN2P3, Institut de Physique Nucl\'eaire, UMR 8608, 91405 Orsay, France}

\date{\today}

\begin{abstract}
The annihilation of proton and antiproton to electron-positron pair, including radiative
corrections due to the emission of virtual and
real photons is considered. The results are generalized to leading and next-to
leading approximations. The relevant distributions are derived and numerical applications are given in the kinematical range accessible to the PANDA experiment at the FAIR facility.
\end{abstract}

\maketitle

% ======================================================================

%----------------------------------------------------
\section{Introduction}
\label{Introduction}
%----------------------------------------------------
The determination of
the proton form-factors in the time-like region is planned at PANDA (FAIR) in
the process of annihilation of proton and antiproton to an electron-positron pair,
for values of antiproton momentum up to 15 GeV/c \cite{epja}. Radiative corrections
(RC) due to the emission of real and virtual photons do affect the measurement of
the experimentally observable quantities, in particular of the differential cross section.
The individual determination of the electric and magnetic proton form factors
requires the precise knowledge of the angular distribution of the final lepton,
in shape and in absolute value. The motivation of this paper to calculate radiative
corrections to this process and to provide the formulae for the differential cross
sections with the adequate accuracy to be used in the experiment.

A lot of attention was paid to the cross-process $e^+e^-\to \mu^+\mu^-(\gamma)$ years ago
(see Ref. \cite{Ku77,Skrzypek:1992vk,Be81}), where the total cross section as well as different distributions were
considered. The purpose of this work is to update and generalize the results obtained in Born approximation and in the lowest order of perturbation theory for the process of radiative annihilation of proton and antiproton to a lepton pair at high energies.

Among the first papers devoted to the annihilation of proton pair to a lepton pair one should  refer to Ref. \cite{Zichichi:1962ni}, where the possibility to
measure the nucleon form factors in the time-like region of momentum transfer was discussed.
In the paper \cite{BR65} the case of polarized particles was considered. More recently, single and double spin polarization observables were calculated in frame of different models and applied to the world data \cite{ETG05} in space-like as well as in time-like region. In Ref. \cite{Ga06} the expression of the cross section and the polarization observables in terms of form factors was extended to the two photon exchange mechanism, in a model independent formalism.

The matrix element of the annihilation process
\be
p(p_+)+\bar{p}(p_-) \to e^+(q_+)+e^-(q_-)
\label{eq:reac}
\ee
in Born approximation has the form:
\ba
M_B=\frac{4\pi\alpha}{q^2}\bar{v}(p_-)\Gamma_{\mu} u(p_+)\bar{u}(q_-)\gamma_{\mu} v(q_+),
\ea
with $q=p_++p_-$, $q^2=s$, $p_\pm^2=M^2$, $q_\pm^2=m^2$, $M(m)$ is the proton(electron) mass, and
\ba
\Gamma_{\mu}(q)=F_1^{str}(s)\gamma_{\mu}+\frac{1}{4M}[\gamma_{\mu},\hat q]F_2^{str}(s),
\ea
where  $F_{1,2}^{str}(s)$ - are the Dirac and Pauli form factors of proton.

For large-angle scattering, the terms of the order of $m^2/M^2$ compared to the ones of order of unity can be neglected ($1+\mathcal O(m^2/M^2))$.

The matrix element squared summed over the spin states is:
\ba
\sum_{spin}|M|^2=16 \cdot (\frac{4\pi\alpha}{s})^2 \cdot \frac{s^2}{4}\{|G_{M}(s)|^2(1+\cos^2\theta)+(1-\beta^2)
|G_E(s)|^2\sin^2\theta\}
\ea
where the magnetic and electric form factors of proton are
\ba
G_E(s)=F_1^{str}(s) + \eta F_2^{str}(s), \quad G_M(s)=F_1^{str}(s)+F_2^{str}(s),
\quad \eta=\frac{s}{4M^2}, \quad \beta^2=1-\frac{4M^2}{s}.
\ea
$M$ is the proton mass, $\beta$ is the proton velocity.

Let us introduce the kinematics invariants
\ba
t&=&(p_- - q_-)^2=-\frac{s}{4}(1+\beta^2-2\beta \cos\theta),\quad
u=(p_+ - q_-)^2=-\frac{s}{4}(1+\beta^2+2\beta \cos\theta)
\nn\\
s&=&\frac{4M^2}{1-\beta^2}, \quad
s+t+u=2M^2,
\ea
where $\theta$ -is the scattering angle between the direction of motion of the antiproton and the electron in the center of mass system (cms).

The differential cross section in Born approximation has the form \cite{Zichichi:1962ni}:
\ba
\frac{d\sigma_B}{dO_-}=\frac{\alpha^2}{4s\beta}\biggl\{|G_M(s)|^2(1+\cos^2\theta)+(1-\beta^2)
|G_E(s)|^2\sin^2\theta\  \biggr\}
\ea
The total cross section is
\ba
\sigma_{tot}=\frac{4\pi\alpha^2}{3\beta s}\biggl[|G_M(s)|^2+\frac{1}{2}|G_E(s)|^2 (1-\beta^2)
\biggr].
\ea

%----------------------------------------------------
\section{Virtual and soft photon emission}
%----------------------------------------------------

Below we will suppose the proton to be point-like particle $G_E(s)=G_M(s)=1$, and RC to the proton and electron vertex can be written in the form:
\begin{gather}
F_{1}(s)=1+\frac{\alpha}{\pi}F_{1}^{(2)}(s)+...\,,\quad
F_{2}(s)=\frac{\alpha}{\pi}F_{2}^{(2)}(s)+...,\quad
F_e(s)=1+\frac{\alpha}{\pi}F_e^{(2)}(s)+...
\end{gather}
Moreover we will consider the case of non-relativistic proton and antiproton,
$\beta \sim 1$.

The Born matrix with one-loop corrections can be written in the form:
\ba
M_{BV}=\frac{1}{1-\Pi(s)}\biggl\{M_{1B}\left [1+\frac{\alpha}{\pi}F_{e}^{(2)}(s)
+\frac{\alpha}{\pi}F_1^{(2)}(s)\right ]+\frac{\alpha}{\pi}M_{2B}F_2^{(2)}(s)\biggr\}
-\left (\frac{\alpha}{\pi}\right )^2M_{box},
\ea
with
\ba
M_{1B}&=&\frac{4\pi\alpha}{s}\bar{v}(p_-)\gamma_{\mu} u(p_+)\bar{u}(q_-)\gamma_{\mu} v(q_+), \nn \\
M_{2B}&=&\frac{4\pi\alpha}{s}\frac{1}{4M}\bar{v}(p_-)(\gamma_\mu \hat{q}-\hat{q}\gamma_\mu)u(p_+)
\bar{u}(q_-)\gamma_\nu v(q_+).
\ea

The real parts of Dirac $F_1^{(2)}(s)$ and Pauli $F_2^{(2)}(s)$ proton form-factors (of QED origin) can be written as:
\cite{Akhiezer,Be81}:
\ba
Re F_1^{(2)}(s)&=&\left (\ln\frac{M}{\lambda}-1\right )\left (1-\frac{1+\beta^2}{2\beta}L_\beta\right)
\nn \\
&+&\frac{1+\beta^2}{2\beta}\left [\frac{1}{3}\pi^2 +\Li\left (\frac{1-\beta}{1+\beta}\right)-
\frac{1}{4}L^2_\beta -L_\beta\ln\left (\frac{2\beta}{1+\beta}\right)\right]-
\frac{1}{4\beta}L_\beta,\nn \\
Re F_2^{(2)}(s)&=&-\frac{1-\beta^2}{4\beta}L_\beta,\quad
L_\beta=\ln\frac{1+\beta}{1-\beta}.
\ea
Only the Dirac form factor is relevant for the lepton:
\ba
Re F_e^{(2)}(s)=\left(\ln\frac{m}{\lambda}-1
\right )(1-L_e)-\frac{1}{4}L_e-\frac{1}{4}L_e^2+2\xi_2,\quad L_e=\ln\frac{s}{m^2}.
\ea
The polarization of vacuum operator $\Pi(s)=\Pi_e(s)+\Pi_{\mu}(s)+\Pi_{\tau}(s)+\Pi_{hadr}(s)$ has a standard form:
\ba
Re \Pi_e(s)&=&\frac{\alpha}{3\pi}\left(L_e-\frac{5}{3}\right), \quad
Im \Pi_e(s)=-\frac{\alpha}{3}, \nn \\
Re \Pi_{\mu}(s)&=&-\frac{\alpha}{\pi}\left[\frac{8}{9}-\frac{\beta_{\mu}^2}{3}-
\beta_{\mu}\left( \frac{1}{2}-\frac{1}{6}\beta_{\mu}^2\right )L_{\mu}\right ],
~
Im \Pi_{\mu}(s)=-\frac{\alpha}{\pi}\cdot \frac{\pi}{2}\left(1-\frac{1}{3}\beta_{\mu}^2\right ), \nn \\
L_{\mu}&=&\ln\frac{1+\beta_{\mu}}{1-\beta_{\mu}}, \quad \beta_{\mu}=\sqrt{1-\frac{4M_{\mu}^2}{s}}.
\ea
The contribution of hadron states to the vacuum polarization is \cite{Eidelman}
\ba
\Pi_{hadr}(s)=-\frac{s}{2\pi^2\alpha}\int\limits_{4M_\pi^2}^\infty
ds'\frac{\sigma_{e^+e^-\to hadr}(s')}{s'-s-i0}.
\ea
In particular for a charged pion pair as hadron state we have
\ba
\sigma_{e^+e^-\to hadr}(s)=\frac{\pi \alpha^2}{3s}\beta_\pi^3|F_\pi(s)|^2,
\quad
\beta_\pi=\sqrt{1-\frac{4M_\pi^2}{s}}.
\ea
with $F_\pi(s)$ is pion form factor. Setting $F_\pi(s)=1$ we have
$$
\Pi_{\pi^+\pi^-}(s)=\frac{2\alpha}{\pi}\left [\frac{1}{12}\ln\frac{1+\beta_{\pi}}{1-\beta_{\pi}}-\frac{2}{3}-2\beta^2_{\pi}-
i\frac{\beta^3_{\pi}}{12}\right ].
$$
The contribution of the box type Feynman diagrams describing the two virtual photons annihilation mechanism is
\ba
M_{box}=\int\frac{d^4k}{i\pi^2}\frac{1}{(k^2-\lambda^2)((q-k)^2-\lambda^2)}
\bar{v}(p_-)\gamma_{\nu}\frac{\hat {p}_+ -\hat {k}+M}{(p_+ -k)^2-M^2} \gamma_{\mu}
u(p_+)
 \nn \\
\times \bar {u}(q_-)
\biggl[\gamma_{\mu}\frac{\hat {q}_- -\hat {k}+m}{(q_- -k)^2-m^2}\gamma_{\nu} +
\gamma_{\nu}\frac{-\hat {q}_+ +\hat {k}+m}{(q_+-k)^2-m^2}\gamma_{\mu}\biggr]v(q_+).
\ea
The interference of Born and box-type amplitudes contribution to the odd part of differential cross section is:
\ba
\label{eq::18}
\frac{d\sigma_{odd}}{d O_-}=\frac{-\alpha^3}{2\pi s^2 \beta}I(t,u,s), \quad
I(t,u,s)=(1-P(t,u))\int\frac{d^4k}{i\pi^2} \cdot \frac{S_e S_p}{(0)(q)(p)(m)},
\ea
where $P(t,u)$ is the exchange operator $P(t,u)f(t,u)=f(u,t)$,
and
\ba
(0)&=&k^2-\lambda^2,~ (q)=(q-k)^2-\lambda^2, ~(p)=k^2-2p_+k+i0,~(m)=k^2-2q_-k+i0,  \nn \\
S_e&=&\frac{1}{4}Tr\hat{q}_+\gamma_\lambda\hat{q}_-\gamma_{\mu}(\hat{q}_--\hat{k})\gamma_\nu, \nn \\
S_p&=&\frac{1}{4}Tr(\hat{p}_++M)\gamma_\lambda(\hat{p}_--M)\gamma_\nu(\hat{p}_+-\hat{k}+M)\gamma_{\mu}.
\label{eq::denom}
\ea

Using the set of one-loop integrals listed in Appendix B we obtain
\ba
I(t,u,s)&=&(u-t)\biggl[\left (\frac{2M^2}{\beta^2}+t+u \right )I_{0qp}-\frac{\pi^2}{6}+\frac{1}{2}L_\beta^2-\frac{1}{\beta^2}L_\beta \biggr]
\nn\\
&+&(2t+s)\biggl[\frac{1}{2}L_{ts}^2-\Li\left (\frac{-t}{M^2-t}\right)\biggr]
-(2u+s)\biggl[\frac{1}{2}L_{us}^2-\Li(\frac{-u}{M^2-u})\biggr]
 \nn \\
&+&(ut-M^2(s+M^2))\biggl[\frac{1}{t}L_{ts}-\frac{1}{u}L_{us}+\frac{u-t}{ut}L_s\biggr]
 \nn \\
&+&\frac{2}{s}\left [ t^2+u^2-4M^2(t+u)+6M^4\right]L_{tu}(L_{M\lambda}+L_s),
\label{eq:20}
\ea
with
\begin{gather}
L_s=\ln\frac{s}{M^2}, \quad L_{tu}=\ln\frac{M^2-t}{M^2-u},
\quad L_{ts}=\ln\frac{M^2-t}{s}, \nn \\
L_\beta=\ln\frac{1+\beta}{1-\beta}, \quad L_{us}=\ln\frac{M^2-u}{s}, \quad L_{M\lambda}=\ln\frac{M^2}{\lambda^2},
\end{gather}
and
\ba
I_{0qp}=\frac{1}{s\beta}\left[L_sL_\beta-\frac{1}{2}L_\beta^2-\frac{\pi^2}{6}+
2\Li\left(\frac{1+\beta}{2}\right)-2\Li\left(\frac{1-\beta}{2}\right)-
2\Li\left(-\frac{1-\beta}{1+\beta}\right)\right].
\ea
For large values of the kinematic invariants $s\sim-t\sim-u\gg M^2$ our result is in agreement with the result previously obtained by I. Kriplovich \cite{Kr73}:
\ba
I(t,u,s)&=&\frac{2}{s}(t^2+u^2)L_{tu}[L_{M\lambda}+L_s]+\frac{1}{2}(t-u)L^2_{tu}+
(t-u)L_{ts}L_{us}+uL_{ts}-tL_{us}\nn\\
&=&s(1+c^2)\left\{\ln\frac{1-c}{1+c}\ln\frac{s}{\lambda^2}+
\frac{1}{2(1+c^2)} \left [c\left (\ln^2\frac{1+c}{2}+\ln^2\frac{1-c}{2}\right )
\right . \right .\nn\\
&&\left .\left .-(1+c)\ln\frac{1-c}{2}+(1-c)\ln\frac{1+c}{2}\right ]\right\}.
\ea
It is useful to note that the coefficient of $L_{M\lambda}$, see Eq. (\ref{eq:20})
$$\frac{2}{s}(t^2+u^2-4M^2(t+u)+6M^4)=s(2-\beta^2\sin^2\theta),$$
is proportional to the Born matrix element squared.

The differential cross section, including Born amplitudes and virtual corrections, becomes:
\ba
\frac{d\sigma_{BV}}{dO_-}=\frac{\alpha^2}{4 s \beta}\biggl\{
\biggl|\frac{1}{1-\Pi}\biggr|^2 \biggl[(2-\beta^2 \sin^2\theta)\left [1+\frac{2\alpha}{\pi}(F_e^{(2)}(s)+F_1^{(2)}(s))\right ]+\nn \\
\frac{4\alpha}{\pi} Re F_2^{(2)}(s)\biggr]-\frac{2\alpha}{\pi s}I(t,u,s)\!\!\biggr\}.
\ea

The soft real photons emission can be taken into account in the standard way:
\ba
\label{eq::25}
\frac{d\sigma^{soft}}{d\sigma_{B}}=-\frac{4\pi \alpha}{(2\pi)^3}\int \frac{d^3k}{2\omega}\left(\frac{p_+}{p_+ k}-
\frac{p_-}{p_- k}+\frac{q_-}{q_- k}-\frac{q_+}{q_+ k} \right)^2 =
\frac{d\sigma^{soft}_{even}}{d\sigma_B}+\frac{d\sigma^{soft}_{odd}}{d\sigma_B},
\ea
where the photon energy in CMS is constrained by: $\omega<\Delta E$,
$\Delta E\ll E=\sqrt{s}/2$. We obtain for the even part:
\ba
\frac{d\sigma^{soft}_{even}}{d\sigma_B}&=&\frac{\alpha}{\pi}
\left \{ -2
\left [\ln\left (\frac{2\Delta E}{\lambda}\right )-
\frac{1}{2\beta}L_{\beta}\right ]-2\ln\left (\frac{\Delta E \cdot m}
{\lambda E}\right )\right .
\nn \\
&+&2 \left . \frac{1+\beta^2}{2\beta}\left[\ln\left(\frac{2 \Delta E}{\lambda}\right)L_{\beta}-
\frac{1}{4}L_{\beta}^2+
\Phi(\beta)\right]+2\left[\ln(\frac{2\Delta E}{\lambda})L_e -\frac{1}{4}L_e^2-\frac{\pi^2}{6}\right]\right \},
\nn
\ea
with $\Phi(\beta)$ :
\ba
\Phi(\beta)&=&\frac{\pi^2}{12}+L_\beta\ln\frac{1+\beta}{2\beta}
+\ln\frac{2}{1+\beta}\ln(1-\beta)+\frac{1}{2}\ln^2(1+\beta)-\frac{1}{2}\ln^22
\nn\\
&-&\Li\left(\beta\right)+\Li\left(-\beta\right)-
\Li\left(\frac{1-\beta}{2}\right), \quad \Phi(1)=-\frac{\pi^2}{6}.
\label{eq:eqA4}
\ea
For the case $s\sim-t\sim-u\gg M^2$ we have \cite{Ku85}
\be
\frac{d\sigma_{BSV}}{d\sigma_B}=\frac{1}{|1-\Pi|^2}D^2_{p\Delta} D^2_{e\Delta}
\left [1+\frac{\alpha}{\pi}(K^p+K^e)_{SV}\right ]
-4\frac{\alpha}{\pi}\ln\frac{\Delta E}{E}\ln\frac{1-c}{1+c}+\frac{\alpha}{\pi}
K^{odd}_{SV}
\ee
with
\ba
D_{p\Delta}&=&1+\frac{\alpha}{2\pi}(L_s-1)\biggl(2\ln\frac{\Delta E}{E}+\frac{3}{2}\biggr), \nn \\
D_{e\Delta}&=&1+\frac{\alpha}{2\pi}(L_e-1)\biggl(2\ln\frac{\Delta E}{E}+\frac{3}{2}\biggr), \nn \\
K_{SV}^p&=&K_{SV}^e=\frac{\pi^2}{3}-\frac{1}{2},
 \nn \\
K_{SV}^{odd}&=&\frac{4}{(1+\cos^2\theta)}
\left[ \cos\theta\left(\ln^2\sin\frac{\theta}{2}+\ln^2\cos\frac{\theta}{2}\right )
+\sin^2\frac{\theta}{2}\ln\cos\frac{\theta}{2}-\cos^2\frac{\theta}{2}
\ln\sin\frac{\theta}{2}\right] \nn \\
&&
-4\ln^2\sin\frac{\theta}{2}+4\ln^2\cos\frac{\theta}{2}
+2\Li\left(\sin^2\frac{\theta}{2}\right)-2\Li\left(\cos^2\frac{\theta}{2}\right).
\ea
Let us note that $K_{SV}^{odd}$ has a different sign than for the annihilation process $ e^++e^-\to \mu^++\mu^-$.

%----------------------------------------------------
\section{Hard photon emission}
%----------------------------------------------------

Let now consider hard  photon  emission contribution:
\be
\bar p(p_-)+p(p_+)\to e^+(q_+)+e^-(q_-)+\gamma(k)
\ee
The differential cross section is the sum of contributions of
emission from leptons and hadrons \cite{Ku77}:
\be
d\sigma=\displaystyle\frac{\alpha^3}{2\pi s}R\displaystyle\frac{dx_-dx dc dz}
{\sqrt{D(c,a_-,z)}}; \,\,D(c,a_-,z)=(z_1-z)(z-z_2), \nn \\
~R=\displaystyle\frac{s}{16(4\pi\alpha)^3}\sum_{spin}|M|^2=R_{even}+R_{odd},
\ee
with
\be
x_{\pm}=\displaystyle\frac{E_{\pm}}{E},~x=\displaystyle\frac{\omega}{E},\nn \\
~z=\cos(\vec{k},\vec{p}_-),~c=\cos(\vec{q}_-,\vec{p}_-), \nn \\
~a_-=\cos(\vec{k},\vec{q}_-),~c_+=\cos(\vec{q}_+,\vec{p}_-),
\ee
and
\be
~z_{1,2}=a_-c\pm \sqrt{(1-a_-^2)(1-c^2)},~ a_-=1-\frac{2(1-x_+)}{xx_-}.
\ee
The energy-momentum conservation gives the relations
\ba
x_++x_-+x=2, \,\,\, x_+c_++x_-c+x z=0.
\ea
The quantity $R$ has been previously calculated in \cite{Ku77,Be81}. For $s\gg M^2$ we have:
\ba
R_{even}&=&-\frac{M^2s}{2s'^2}\left [\displaystyle\frac{t^2+u^2}{(p_-k)^2}+ \displaystyle\frac{t'^2+u'^2}{(p_+k)^2}\right ]
-\displaystyle\frac{m^2}{2s}\left [\displaystyle\frac{t^2+u'^2}{(q_-k)^2}+ \displaystyle\frac{t'^2+u^2}{(q_+k)^2}\right ]\nn\\
&&
+\displaystyle\frac{t^2+t'^2+u^2+u'^2}{4s'}\left (\displaystyle\frac{s}{p_+k p_-k}+ \displaystyle\frac{s'}{q_+k q_-k}\right );\nn\\
R_{odd}&=&\frac{t^2+t'^2+u^2+u'^2}{4s'}\left (\displaystyle\frac{t}{p_+k\cdot q_+k}+
\displaystyle\frac{t'}{p_-k\cdot q_-k}-
\displaystyle\frac{u'}{p_-k\cdot q_+k}-
\displaystyle\frac{u}{p_+k\cdot q_-k}
\right ),\nn
\ea
with
\ba
t&=&-2p_+q_+,~u=-2p_+q_-,~s=2p_+p_-,~
t'=-2p_-q_-,~u'=-2p_-q_+,~s'=2q_+q_-,\nn\\
&&t+t'+u+u'+s+s'=0.
\nn
\ea

Performing the simplifications and omitting the contributions of terms of order
\ba
O\left (\frac{M^2}{s}\right )
\ea
compared to the ones which lead to contributions of the order of unity,
we obtain
\ba
R_{even}&=&R_M+R_p+R_q, \nn \\
R_M&=&-(1+P_c)\frac{2M^2}{s\bar{x}^2}\frac{x_-^2}{x^2}\frac{\sigma_+}{(1-\beta z)^2}-
\frac{m^2}{s}(1+c^2)\biggl[\frac{1}{\bar{x}_+^2}+\frac{1}{\bar{x}_-^2}\biggr]; \nn \\
R_p&=&2(1+P_c)\frac{X}{\bar{x}x^2(1-\beta z)};\,\,\,
R_q=\frac{X}{\bar{x}_+\bar{x}_-}, \qquad
\ea
where $\sigma_+=\bar{x}^2(1-c)^2+(1+c)^2$, $\bar{x}=1-x,\bar{x}_\pm=1-x_\pm$ and
the exchange operator $P_c$ is defined as $P_cf(c)=f(-c)$, (see Eq. (\ref{eq::18}), $P_c=P(t,u)$).
The quantity
\ba
X=\frac{t^2+t'^2+u^2+u'^2}{s^2}
\ea
can be written in two (equivalent) forms. In the case when it is convoluted
with the "singular" factor $1/(1-\beta z)$, keeping in mind that one has to further
integrate on $x_-$ and $z$ (see details in Appendix A) it has the form:
\ba
X&=&A_1x_-^2+A_2(1-z)^2+A_3x_-(1-z)(c-t)+ \nn \\
&&A_4x_-^2(c-t)^2+A_5x_-(1-z)+A_6x_-^2(c-t),
\ea
with $x_-=2\bar{x}/r, r=2-x(1-t)$. Here and further we use the notation $t=a_-=1-2\bar{x}_+/(xx_-)$. The coefficients $A_i$ are
\ba
A_1&=&\frac{1+\bar{x}^2}{4\bar{x}^2}\sigma_+~; A_2=\frac{1}{2}x^2;~A_3=\frac{x^3}{2\bar{x}}; \nn \\
A_4&=&\frac{x^2(1+\bar{x}^2)}{4\bar{x}^2};~ A_5=\frac{x}{2\bar{x}}[\bar{x}^2(1-c)-(1+c)];~
A_6=-\frac{x}{2\bar{x}^2}[\bar{x}^3(1-c)+(1+c)].
\label{eq:eqai}
\ea

For "nonsingular" terms entering in $R_q$ we have (following from the definition):
\ba
X&=&\frac{1}{2}[x_+^2(1+c_+^2)+x_-^2(1+c^2)]= \nn \\
&&\frac{1}{2}[(2-x_--x)^2+(x_-c+zx)^2+x_-^2(1+c^2)].
\ea
The contribution which is odd under the action of $P_c$ has the form (in agreement with Ref. \cite{Ku77})
\ba
R_{odd}&=&\frac{1}{\bar{x}}\biggl[\frac{2}{x^2(1-\beta^2z^2)}[xx_+(c_+-z)+xx_-(z-c)]+
\frac{\bar{x}}{\bar{x}_+\bar{x}_-}[x_-\bar{x}_-c-x_+\bar{x}_+c_+]+2(x_-c-x_+c_+) \nn \\
&& +\frac{x_+(1+c_+)}{2x\bar{x}_-(1+\beta z)}[2x_+x_-(1-c c_+)+xx_-(1-z c)-x_+\bar{x}_+(1-c_+)-xx_-(1-c)]-\nn \\
&&\frac{x_+(1-c_+)}{2x\bar{x}_-(1-\beta z)}[2x_+x_-(1-c c_+)+xx_-(1-z c)-x_+\bar{x}_+(1+c_+)-xx_-(1+c)]+ \nn \\
&&\frac{x_-(1-c)}{2x\bar{x}_+(1-\beta z)}[2x_+x_-(1-c c_+)+xx_+(1-z c_+)-x_+\bar{x}_+(1+c_+)-xx_-(1+c)]- \nn \\
&&\frac{x_-(1+c)}{2x\bar{x}_+(1+\beta z)}[2x_+x_-(1-c c_+)+xx_+(1-z c_+)-x_+\bar{x}_+(1-c_+)-xx_-(1-c)]\biggr]. \qquad
\ea
Let now focus on the distribution $d\sigma/(d x d c)$.

The integration on the two other variables ($x_-,z$) (see Appendix \ref{AppendixA}) leads to
\ba
\int\frac{dx_- dz}{\pi \sqrt{D}}R_M=-\frac{8\bar{x}}{x}(1+P_c)
\frac{\sigma_+}{A^4}-\frac{2\bar{x}}{x}(1+c^2), \quad
A=2-x(1-c).
\ea
Using the relations
\ba
\frac{1}{\bar{x}_+\bar{x}_-}=\frac{1}{x}\left[\frac{1}{\bar{x}_+}+\frac{1}{\bar{x}_-}\right],
\quad
x_+c_++x_-c+z x=0,
\ea
and performing the integration over $z$, the expression for $R_q$ can be written as
\ba
R_q=\frac{1+c^2}{2x} \left [(1+\bar{x}^2)\left (\frac{1}{\bar{x}_+}+\frac{1}{\bar{x}_-}\right )+\frac{2x\bar{x}}{x_-^2}-2x\right].
\ea
The integration on $x_-$ leads to
\ba
\int\frac{dx_- dz}{\pi \sqrt{D}}R_q=\frac{1+\bar{x}^2}{x}(1+c^2)\ln\frac{s\bar x}{m^2}
\ea
and
\ba
\int\frac{dx_- dz}{\pi \sqrt{D}}R_p&=&(1+P_c)\frac{4}{x}\biggl[4\bar{x}^2A_1J_1+A_2J_2+ \nn \\
&&2\bar{x}A_3J_3+4\bar{x}^2A_4J_4+2\bar{x}A_5J_5+4\bar{x}^2A_6J_6 \biggr],
\ea
with $A_i$ given above and $J_i$ listed in Appendix A. Note that
only quantity $J_1$ contains the logarithmically-enhanced contribution
\ba
J_1=\frac{1}{A^4}\ln\frac{s}{M^2}+\Delta J_1. \nn
\ea
Moreover, the coefficients $A_2-A_6$ as well as the quantity
$\Delta J_1$ are proportional to $x$ at small $x$.

Performing the integration over $z$ we obtain
\ba
\frac{d\sigma_{odd}^{hard}}{d x d c}&=&\frac{\alpha^3}{s}(1-P_c)\int\limits_{\bar{x}-\frac{m^2x}{s\bar{x}}}^{1-\frac{m^2x}{s\bar{x}}}
\frac{d x_-}{x\bar{x}}
\left\{c\left (2x\bar{x}_+-\frac{x}{\bar{x}_+}+\frac{x\bar{x}^2}{\bar{x}_-}\right ) \right.
\nn\\
&+&\frac{1}{\sqrt{R}}\left [\frac{1+c}{\bar{x}_+}(c^2+c x+1-x+x^2)
-\frac{1-c}{\bar{x}_-}(c^2\bar{x}^2+c x\bar{x}+1-x+x^2)\right .
\nn \\
&-& x A(1+c^2)+x^2c(1-c)^2-2x(c^3-c^2+c+1)+2(1-c^2)\Big] \Big \},
\ea
with $R=(c-t)^2+(1-\beta^2)(1-c^2)$.

Performing the integration over $x_-$ we obtain
\ba
\frac{d\sigma_{odd}}{d x d c}&=&\frac{\alpha^3}{2s}(1-P_c)\frac{1}{A}
\left\{\left [c^2+x c+1-x+x^2][\frac{4}{x}\ln\frac{1+c}{2}+
\frac{1+c}{\bar{x}}\ln\frac{s\bar{x}(1+c)^2}{m^2A^2}\right ] \right.\nn \\
&&+\left [-\frac{4}{x}\ln\frac{\bar{x}(1-c)}{2}+(1-c)\ln\frac{s\bar{x}^3(1-c)^2}{m^2A^2}\right ]\left[c^2\bar{x}^2+c x\bar{x}+1-x+x^2\right ]+ \nn \\
&&2c(2c+x(1-c))\ln\frac{4s\bar{x}}{m^2A^2}+2(1+c^2)\left[\frac{6\bar{x}}{A}-(2-x)-\frac{A}{4\bar{x}}(1+\bar{x}^2)\right] +\nn \\
&&\left .\left(-\frac{4}{A}+\frac{2-x}{\bar{x}}\right)\left[x^2c(1-c)^2-2x(c^3-c^2+c+1)+2(1-c^2)\right]\right\}.
\ea
Performing the integration on the photon energy fraction $\Delta=\frac{\Delta E}{E}<x< 1-\frac{4m^2}{s}$
we obtain (see Appendix A)
\ba
\frac{d\sigma_{odd}^{hard}}{dc}&=&\frac{\alpha^3}{2s}\biggl\{4(1+c^2)\ln\frac{1+c}{1-c}\ln\frac{1}{\Delta}+
c\left[ -6-\frac{\pi^2}{3}-\ln^2\left(\frac{1-c}{1+c}\right)\right]+2(1-3c^2)\ln\frac{1+c}{1-c}+ \nn \\
&&
\frac{6}{1-c}\left(-1+2\frac{\ln\frac{2}{1+c}}{1-c}\right)-\frac{6}{1+c}
\left(-1+2\frac{\ln\frac{2}{1-c}}{1+c}\right)+ \nn \\
&&
2(1+c^2)\biggl[\int\limits_{\frac{1-c}{2}}^{\frac{1+c}{2}}
\frac{dx}{x}\ln(1-x)+\frac{1}{2}\ln^2\left (\frac{1-c}{2}\right )-\frac{1}{2}\ln^2\left (\frac{1+c}{2}\right )\biggr]\biggr\}.
\label{eq::46}
\ea

The odd part of differential distribution does not have logarithmic enhancement.

The logarithmically - enhanced contribution from hard
photon emission to the even part of differential cross section has the form:
\ba
\frac{d\sigma^{hard}_{even}}{dx dc} =
\frac{\alpha}{2\pi}\cdot \biggl\{\biggl[\frac{d\sigma_B(\bar{x} p_-, p_+)}{d c}+
\frac{d\sigma_B(p_-,\bar{x} p_+)}{d c}\biggr] \cdot
\frac{1+\bar{x}^2}{x}\biggl(\ln\frac{s}{ M^2}-1\biggr) +\nn \\
2\frac{d\sigma_B(p_-,p_+)}{d c}\biggl[\frac{1+\bar{x}^2}{x}\ln\left (\frac{s}{m^2}-1\right )\biggr]+
\frac{\pi\alpha^2}{s}K(x,c)\biggr\},
\ea
with
\ba
K(x,c)&=&(1+c^2)\left (x+\frac{1+\bar{x}^2}{x}\ln\bar{x}\right )+(1+P_c)\left [\frac{4x\sigma_+}{A^4}+\frac{4}{x}Q(x,c)\right ], \nn \\
Q(x,c)&=&(1+\bar{x}^2)\Delta J_1\sigma_++A_2J_2+2\bar{x}A_3J_3+4\bar{x}^2A_4J_4+2\bar{x}A_5J_5+4\bar{x}^2A_6J_6.
\ea

%----------------------------------------------------
\section{Discussion}
%----------------------------------------------------

Extracting the logarithmically - enhanced terms in virtual and soft photon (27) and hard photon (47)
emission and keeping in mind the factorization theorem \cite{Ku85} we can express the differential cross section for the process
\be
\bar p(p_-)+p(p_+)\to e_+(y_+)+e_-(y_-)+(\gamma(k))
\ee
in the form
\ba
\displaystyle\frac{d\sigma}{dc dy_+ dy_-}&=&
\int dx_+dx_-{\cal D}(x_+,L_s){\cal D}(x_-,L_s)
\displaystyle\frac{d\sigma_B(x_-p_-,x_+p_+,z_+,z_-)}{d c} \nn \\
&&\frac{1}{|\Pi(sx_+x_-)|^2}\left (1+\displaystyle\frac{\alpha}{\pi} K\right )
\displaystyle\frac{1}{z_+z_-}{\cal D}\left (\displaystyle\frac{ y_+}{z_+},L_e\right )
{\cal D}\left (\displaystyle\frac{y_-}{z_-},L_e\right )+ \nn \\
&&+\left( \displaystyle\frac{d\sigma}{d c}\right )^{odd},
\ea
where $y_{\pm}$ is the energy fraction of the final lepton. The non-singlet structure function ${\cal D}$ of the leptons is \cite{Ku85,Skrzypek:1992vk}
\ba
{\cal D}(x,L)&=& \delta(x-1)+ \frac{b}{4} P^{(1)}(x)+\frac{1}{2!}(\frac{b}{4})^2 P^{(2)}(x)+...,\nn\\
P^{(1)}(x) &=& \left(\frac{1+x^2}{1-x}\right)_+,
\qquad
P^{(2)}(x) =P \otimes P(x),
\ea
or equivalently
\ba
{\cal D}(x,L)=\frac{1}{2}b(1-x)^{\frac{b}{2}-1}\left (1+\frac{3 b}{8}\right )-\frac{1}{4}b(1+x)+O(b^2),
\quad
b=\frac{2\alpha}{\pi}(L-1).
\ea

\begin{figure}
\includegraphics[width=0.7\textwidth]{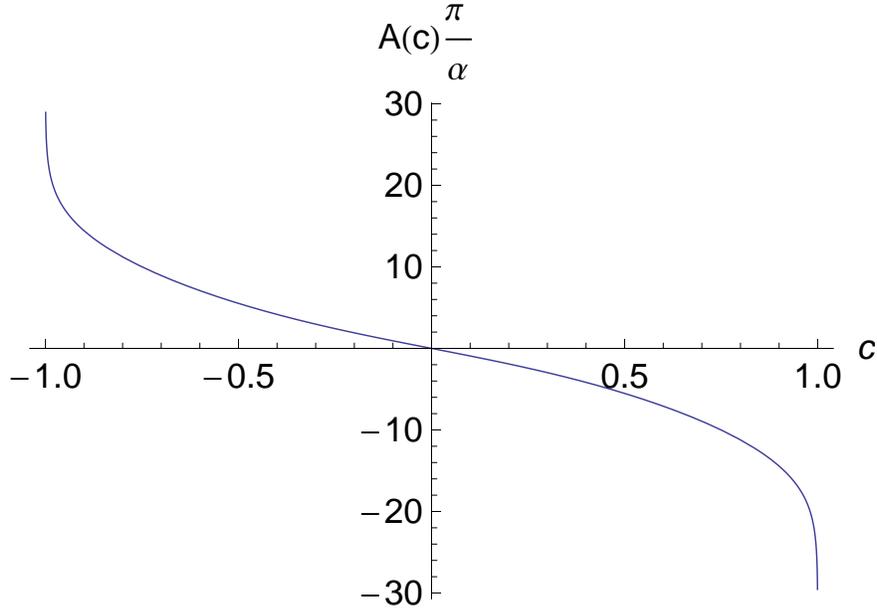}
\caption{Asymmetry  A(c) $\cdot \frac{\pi}{\alpha}$ ( see(\ref{eq::56}))
as a function of $c, \,\,\,c=\cos\widehat{\vec p_{-} \vec q_{-}}$.}
\label{Fig1}
\end{figure}

The shifted Born cross section for the hard subprocess is:
\be
\frac{d\sigma_B(x_-p_-,x_+p_+,z_+,z_-)}{d c}=\displaystyle\frac{d\sigma_B(x_-,x_+ )}{d c}=
\displaystyle\frac{4\pi\alpha^2}{s}
\displaystyle\frac{x_-^2(1-c)^2+x_+^2(1+c)^2}{[x_-(1-c)+x_+(1+c)]^4},
\ee
and the energy fraction of the leptons in the hard subprocess is:
\be
z_-=\displaystyle\frac{x_-^2(1-c)+x_+^2(1+c)}{x_-(1-c)+x_+(1+c)};~
z_+=\displaystyle\frac{2x_- x_+}{x_-(1-c)+x_+(1+c)};~c=\cos\widehat{\vec p_-\cdot\vec q_-}.
\ee
The explicit form of $(d\sigma/d c)^{odd}$ is the sum of soft, virtual and hard photon
emission contributions ( see (\ref{eq::18},\ref{eq::25},\ref{eq::46}):
\ba
\frac{d \sigma^{odd}}{d c}& =& \frac{\alpha^3}{2 s}F(c), \nn \\
F(c)&=&c\left(-6-\frac{\pi^2}{3}+2 \ln\frac{2}{1+c}
\ln\frac{2}{1-c}+\ln\frac{4}{1-c^2}\right )+3(1-2c^2)\ln\frac{1+c}{1-c}+ \nn \\
&&\frac{6}{1-c}\left(-1+\frac{2}{1-c}\ln\frac{2}{1+c}\right )
-\frac{6}{1+c}\left(-1+\frac{2}{1+c}\ln\frac{2}{1-c}\right )+ \nn \\
&&4(1+c^2)\left[Li_2\left (\frac{1-c}{2}\right )-Li_2\left(\frac{1+c}{2}\right)\right ].
\ea
The relevant charge-asymmetry
\ba
\label{eq::56}
A(c)=\frac{d\sigma(c)-d\sigma(-c)}{d\sigma(c)+d\sigma(-c)}=\frac{\alpha}{\pi}\frac{F(c)}{1+c^2},
\ea
turns out to be rather large to be measured in experiment: $|A|\sim 4-5\%$. It is presented
in Fig. \ref{Fig1}. We note the numerical agreement with the calculation of Ref. \cite{Ku77}.

\begin{figure}
\includegraphics[width=0.7\textwidth]{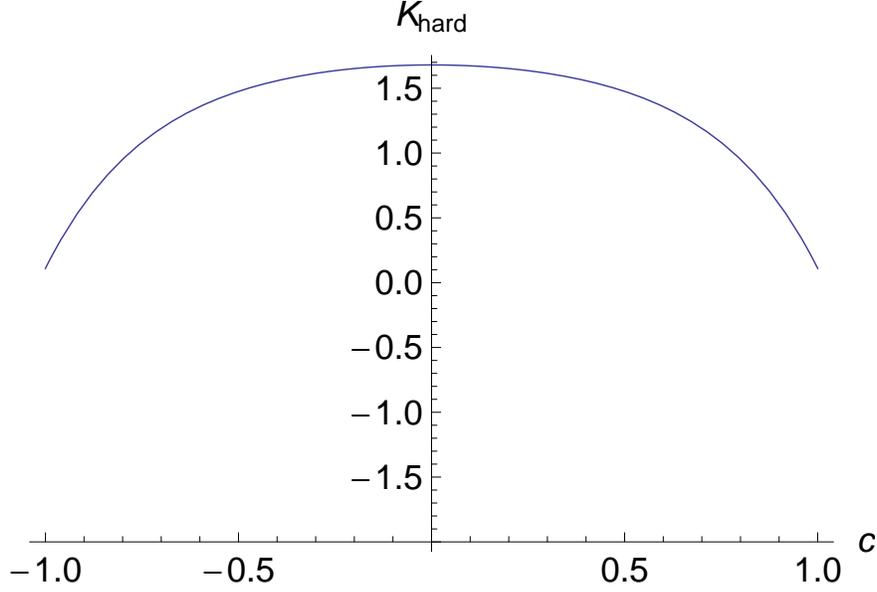}
\caption{K-factor for even part of cross-section  $K_{hard}(c)$ (see (\ref{eq::59}))
as a function  of $c, \,\,\,c=\cos\widehat{\vec p_{-} \vec q_{-}}$.}
\label{Fig2}
\end{figure}

The non - leading terms (which do not contain the "large logarithms" $L_e, L$) are contained in the factor
$(1+\frac{\alpha}{\pi}K)$ with $K=K_{SV}+K_{hard}(x_+,x_-,c), \,\, K_{SV}=\frac{2}{3}\pi^2-1$ and
\ba
K_{hard}(x_-,x_+,c)&=&\frac{[x_-(1-c)+x_+(1+c)]^4}{8[(x_-^2(1-c)^2+x_+^2(1+c)^2]}F(x_-,x_+,c); \nn \\
F(x_-,x_+,c)&=&\frac{1}{2}(1+c^2)[\bar{x}_-+\frac{1+x_-^2}{\bar{x}_-}\ln x_-+\bar{x}_++\frac{1+x_+^2}{\bar{x}_+}\ln x_+] \nn \\
&&+\frac{4r_-\bar{x}_-}{A_-^4}+\frac{4}{\bar{x}_-}Q(\bar{x}_-,c)+\frac{4r_+\bar{x}_+}{A_+^4}+
\frac{4}{\bar{x}_+}Q(\bar{x}_+,-c),
\ea
with
\ba
r_-=x_-^2(1-c)^2+(1+c)^2,\qquad A_-=1+x_-+c(1-x_-); \nn \\
r_+=(1-c)^2+x_+^2(1+c)^2,\qquad A_+=1+x_+-c(1-x_+).
\ea
The quantity
\ba
\label{eq::59}
K_{hard}\left (\frac{1}{2},\frac{1}{2},c\right )=\frac{1}{8(1+c^2)}F\left (\frac{1}{2},\frac{1}{2},c\right )
\ea
is presented in Fig.\ref{Fig2}.

In the application of the present calculation to the kinematical range covered by PANDA ($s=15$ GeV$^2$) we can neglect the corrections related
with neutral weak currents (annihilation through neutral $Z$-boson). Indeed, the relevant
correction $(1+\delta_W)$ is very close to unity $\delta_W\sim a_v s/M_Z^2<10^{-3}$.

The distributions on the energy fractions of the muons (Dalitz-plot distribution),
the inclusive distribution and the total cross-section
of the process $e^+e^-\to \mu^+\mu^-\gamma$ was considered
in the papers \cite{Ku77},\cite{Be81}. The relevant results for the process
considered here
can be obtained by a simple replacement.

\appendix
%--------------------------------------------------------------------
\section{Integrals for the case of hard photon emission}
\label{AppendixA}
%--------------------------------------------------------------------

Transforming the phase volume of final 3-body state as
\ba
\frac{1}{s\pi^2}\frac{d^3q_-}{2E_-}\frac{d^3q_+}{2E_+}\frac{d^3k}{2\omega}\delta^4(p_++p_--q_+-q_--k)&=& \nn \\
\frac{1}{s\pi^2}\frac{d^3q_-}{2E_-}\frac{d^3k}{2\omega}\delta((p_++p_--q_--k)^2-m^2) &=& \nn \\
\frac{1}{32\pi}x_- x dx_- d x d c dO_\gamma \delta[(1-x_--x+\frac{1}{2}x x_-(1-t)]&=& \nn \\
\frac{d c d x}{16\pi}d x_-\frac{2 d z}{\sqrt{D}}=
\frac{d c d x}{4\pi}\frac{x\bar{x}}{r^2}\frac{d t d z}{\sqrt{D}},&&
\ea
where $D=D(z,c,t)=1-z^2-t^2-c^2+2z c t=(z-z_1)(z_2-z)$, $c=\cos(\vec{p}_-,\vec{q}_-), z=\cos(\vec{k},\vec{p}_-)$
-are the cosines of
electron and photon emission angles, counting from the antiproton direction,
\ba
x&=&\frac{2\omega}{\sqrt{s}},\,\,x_\pm=\frac{2 E_\pm}{\sqrt{s}},
\qquad x_-=\frac{2\bar x}{r},~x_++x_-+x=2, \nn \\
t&=&1-\frac{2(1-x_+)}{x x_-}=\cos(\vec{k},\vec{q}_-),\,\, r=2-x(1-t).
\ea
Let us introduce the new 4-vector
$\tilde{k}=k-xp_-$, in order to calculate $X$ in the limit $z\to 1$:
\ba
4X&=&(x_-(1-c))^2+(x_-(1+c))^2+(-\bar{x}x_-(1-c)-x(1-z)+xx_-(c-t))^2+ \nn \\
&&\frac{1}{\bar{x}^2}(-x_-(1+c)+x\bar{x}(1-z)+xx_-(c-t))^2= \nn \\
&&4[A_1x_-^2+A_2(1-z)^2+A_3x_-(1-z)(c-t)+A_4x_-^2(c-t)^2+ \nn \\
&& A_5x_-(1-z)+A_6x_-^2(c-t)],
\ea
with $A_i$ given in Eq. \ref{eq:eqai}.

Here we used the conservation law in the form $p_+ +\bar{x}p_- =q_++q_-+\tilde{k}$
and the kinematic relations
\ba
\tilde{k}^2&=&-x\chi_-, \,\,2p_-\tilde{k}=-2p_+\tilde{k}=\chi_-; \,\, \chi_-=2p_-k=\frac{s}{2}x(1-z); \nn \\
&& 2q_-\tilde{k}=\frac{s}{2}xx_-(c-t).
\ea

The relevant integrals on photon emission angle $\theta, z=\cos\theta$ are
\ba
\frac{1}{\pi}\int\limits_{z_1}^{z_2}\frac{d z}{\sqrt{D(c,t,z)}}[\frac{1}{(1-\beta z)^2};
\frac{1}{1-\beta z};1, z, z^2]&=&
\biggl[\frac{1-ct}{Q^3};\frac{1}{Q};1; c t;\frac{1}{2}(1-c^2-t^2+3c^2t^2)\biggr],
\nn \\
Q&=&\sqrt{(c-t)^2+(1-\beta^2)(1-c^2)},
\ea
where $\beta$ is implied to be a positive and real quantity not exceeding unity, with $1-\beta^2<<1$.
For $\beta=1$ we obtain $Q=|c-t|$. \\

The domains of variation of the lepton energy fractions are deduced from the kinematic relation
$$2qk_-=\frac{s}{2}x x_-(1-\beta_-t)=s(1-x_+)> \frac{s}{4}x x_-(1-\beta^2_-).$$
We obtain
\ba
\bar{x}-\frac{m^2x}{s\bar{x}}<x_\pm< 1-\frac{m^2x}{s\bar{x}},
\ea
and
\ba
t_-=-1+\frac{2m^2}{s\bar{x}}<t<1-\frac{2m^2}{s\bar{x}^2}=t_+.
\ea
The integrals on electron energy fraction (or on the variable $t$) associated with the photon emission by the initial nucleons are
\ba
J_1&=&\int\limits_{-1}^1\frac{d t}{r^4R}= \frac{1}{A^4}\ln\frac{s}{M^2}+\Delta J_1; \nn \\
\Delta J_1 &=&\frac{1}{A^4}\biggl[\ln\frac{A^2}{4\bar x}+\frac{x(1+c)}{24\bar{x}^3}(A^2+5\bar{x}A+22)-
\frac{x(1-c)}{24}(A^2+5A+22)\biggr]; \nn \\
J_2&=&\int\limits_{-1}^1\frac{d t}{r^2}(1-ct)=\frac{1}{x^2}\biggl[\frac{x}{2\bar{x}}(x+c(2-x))-c\ln\frac{1}{\bar{x}}\biggr]; \nn \\
J_3&=&\int\limits_{-1}^1\frac{d t}{r^3}(c-t)=\frac{x+c(2-x)}{8\bar{x}^2}; \nn \\
J_4&=&\int\limits_{-1}^1\frac{d t}{r^4}\frac{(c-t)^2}{|c-t|}=\frac{1}{24A^2}\biggl[\frac{(1+c)^2}{\bar{x}^3}(3-x(2-c))+
(1-c)^2(3-x(1-c))\biggr]; \nn \\
J_5&=&\int\limits_{-1}^1\frac{d t}{r^3}=\frac{2-x}{8\bar{x}^2}; \nn \\
J_6&=&\int\limits_{-1}^1\frac{d t}{r^4}\frac{c-t}{|c-t|}=\frac{1}{24A^3}\biggl[\frac{1+c}{\bar{x}^3}
(A^2+2\bar{x}A+4\bar{x}^2)-(1-c)(A^2+2A+4)\biggr], \nn \\
A&=&2-x(1-c).
\ea
For the odd part of cross section we use ($r=2-x(1-t)$, $A=2-x(1-c)$)
\be
\int\limits_{-1}^1 d t\biggl[\displaystyle\frac{1}{r};\displaystyle\frac{1}{r^2};\displaystyle\frac{1}{r^3}\biggr]=
\biggl[\displaystyle\frac{1}{x}\ln\displaystyle\frac{1}{\bar{x}};
\displaystyle\frac{1}{2\bar{x}};\displaystyle\frac{2-x}{8\bar{x}^2}\biggr];
\ee
\ba
\int\limits_{-1}^1 dt\frac{c-t}{|c-t|}\biggl[\frac{1}{r};\frac{1}{r^2};\frac{1}{r^3}\biggr]&=&
\biggl[\frac{1}{x}\ln\frac{A^2}{4\bar{x}};\frac{1}{2A}\biggl[\frac{1+c}{\bar{x}}-(1-c)\biggr]; \nn \\
&&\frac{1}{8A^2}\biggl(\frac{(1+c)(A+2\bar{x})}{\bar{x}^2}-(1-c)(A+2)\biggr)\biggr]; \nn
\ea
In order to perform the integration over the photon energy fraction, we use
\ba
\int\limits_0^1 dx \biggl[\frac{1}{A}; \frac{1}{A^2}; \frac{\ln A}{A}; \ln\bar{x} \biggr]&=&\biggl[\frac{1}{1-c}\ln\frac{2}{1+c}; \frac{1}{2(1+c)}; \frac{1}{2(1-c)}(\ln^22-\ln^2(1+c));
\frac{1}{1-c}Li_2(-\frac{1-c}{1+c})\biggr],
\nn \\
\int\limits_0^{1-\frac{4m^2}{s}}\frac{dx}{\bar x}\biggl[\ln\bar {x}; \ln A\biggr]
&=&\biggl[-\frac{1}{2}\ln^2\frac{s}{4m^2};
\ln\frac{s}{4m^2}\ln(1+c)-Li_2(-\frac{1-c}{1+c})\biggr].
\ea

%--------------------------------------------------------------------
\section{One loop momentum integrals}
\label{AppendixB}
%--------------------------------------------------------------------

\begin{gather}
I_{ab}=\int\frac{d^4k}{i\pi^2 (a)(b)},\quad I_{abc}=\int\frac{d^4k}{i\pi^2(a)(b)(c)},\quad
I_{0qpm}=\int\frac{d^4k}{i\pi^2(0)(q)(p)(m)},\nn\\
J_{abc}^{\mu} =\int\frac{d^4k \cdot k_{\mu}}{i\pi^2 (a)(b)(c)}.
\label{eq:eqB1}
\end{gather}
The standard procedure for joining the denominators leads to (for designations of the
denominators see eq. (\ref{eq::denom})):
\ba
Re I_{0q}&=&L_{\Lambda}-L_{s}+1,~L_{\Lambda}=\ln\frac{\Lambda^2}{M^2},\nn  \\
I_{0p}&=&I_{qp}=L_{\Lambda}+1, ~L=\ln\frac{M^2}{m^2}, \nn \\
I_{0m}&=&I_{qm}=L_{\Lambda}+L+1,\nn \\
I_{pm}&=&L_{\Lambda}+1+\left(\frac{M^2}{u}-1\right)\ln\left(1-\frac{u}{M^2}\right), \nn
\ea
with $\Lambda$ is the ultraviolet cut-off parameter.
The real parts of three and four denominators scalar integrals are:

\ba I_{0qp}&=&\int\limits_0^1\frac{dx}{xs+M^2(1-x)^2}\ln\frac{sx}{M^2(1-x)^2},
~
Re I_{0qp}=\frac{1}{s}\biggl(\frac{1}{2}\ln^2\frac{s}{M^2}+\frac{\pi^2}{6}\xi_2\biggr),~s \gg M^2,
\nn \\
Re I_{0qm}&=&\frac{1}{s}\biggl[\frac{1}{2}\ln^2\frac{s}{m^2}+\frac{\pi^2}{6}\xi_2\biggr],
\nn \\
I_{0pm}&=&\frac{-1}{2(M^2-u)}\{\ln\frac{mM}{\lambda^2}\ln\frac{(M^2-u)^2}{m^2M^2}-2Li_2(\frac{-u}{M^2-u})+ \nn \\
&&\ln^2(\frac{M^2-u}{M^2})+\ln\frac{M^2-u}{M^2} \cdot L \};
\label{eq:eqB2}\\
I_{0pm}&=&I_{qpm}=\frac{-1}{2(M^2-u)}\{(\ln\frac{M^2}{\lambda^2}-\frac{1}{2}L)(2\ln\frac{M^2-u}{M^2}+L)-
2Li_2(\frac{-u}{M^2-u})+ \nn \\
&&\ln^2(\frac{M^2-u}{M^2})+L\cdot \ln\frac{M^2-u}{M^2}\},
\label{eq:eqB3} \\
I_{0pm}&=&\frac{-1}{2(M^2-u)}\{L_{\lambda}(2\ln\frac{M^2-u}{M^2}+L)-\frac{1}{2}L^2+
\ln^2(\frac{M^2-u}{M^2})-
2Li_2(\frac{-u}{M^2-u})\}, \label{eq:eqB4} \\
I_{0pqm}&=&-\frac{1}{s(M^2-u)}\ln\frac{s}{\lambda^2}.\label{eq:eqB5}
\ea
The relevant vector integrals are:
\ba
J_{0qp}^{\mu}&=&\alpha_{0qp}p_{+}^{\mu}+\beta_{0qp}q^{\mu},~
\alpha_{0qp}=\frac{1}{s\beta^2}(sI_{0qp}-2L_s), \,\,\beta_{0qp}=-\frac{1}{s\beta^2}(2M^2 I_{0qp}-L_s),
\nn\\
J_{0qm}^{\mu}&=&\alpha_{0qm}q^{\mu}+\beta_{0qm}q_{-}^{\mu},~
\alpha_{0qm}=\frac{1}{s}(1+L_s); ~\beta_{0qm}=-\frac{2}{s}(L_s+L)+I_{0qm},
\nn\\
J_{0pm}^{\mu}&=&\alpha_{0pm}p_{+}^{\mu}+\beta_{0pm}q_{-}^{\mu},~
\alpha_{0pm}=\frac{1}{M^2-u}[I_{pm}-I_{0p}],\nn\\
\beta_{0pm}&=&\frac{1}{M^2-u}[I_{pm}-I_{0m}-2M^2 \alpha_{0pm}],\nn\\
J_{qpm}^{\mu}&=&I_{qpm}q^{\mu}+\alpha_{qpm}p_{-}^{\mu},~
\alpha_{qpm}=\frac{I_{qp}-I_{pm}}{M^2-u}, \nn \\
\beta_{qpm}&=&\frac{(M^2+u)(I_{pm}-I_{qm})+2M^2(I_{qm}-I_{qp})}{(M^2-u)^2}.
\ea

%\begin{figure}
%\includegraphics[width=0.5\textwidth]{Fig1.eps}
%\caption{}
%\label{Fig1}
%\end{figure}

\end{document}